\documentclass[10pt,emulateapj,apj]{emulateapj}
\shorttitle{Large-Scale Structure as a Cosmic Ruler}
\shortauthors{Park \& Kim}
\begin{document}
\title{Large-Scale Structure of the Universe as a Cosmic Standard Ruler}
\author{Changbom Park and Young-Rae Kim}
\affil{School of Physics, Korea Institute for Advanced Study, Seoul 130-722, Korea}
\email{cbp@kias.re.kr, yrk@kias.re.kr}

\begin{abstract}
We propose to use the large-scale structure of the universe as a cosmic standard 
ruler, based on the fact that the pattern of galaxy distribution 
should be maintained in the course of time on large scales.
By examining the scale-dependence of the pattern in different redshift intervals
it is possible to reconstruct the expansion history of the universe, 
and thus to measure the cosmological parameters governing the expansion of the universe.
The features in the galaxy distribution that can be used as standard
rulers include the topology of large-scale structure and the overall shapes of 
galaxy power spectrum and correlation function.  
The genus, being an intrinsic topology measure, is resistant against the non-linear
gravitational evolution, galaxy biasing, and redshift-space distortion effects,
and thus is ideal for quantifying the primordial topology of the large-scale structure.
The expansion history of the universe can be constrained by comparing 
among the genus measured at different redshifts.
In the case of initially Gaussian fluctuations the genus accurately recovers the slope 
of the primordial power spectrum near the smoothing scale, and the expansion
history can be constrained by comparing between the predicted and 
measured genus.
\end{abstract}


\section{Introduction}

There are three kinds of phenomena of the universe that are currently 
used to constrain cosmological models. 
The first is the primordial fluctuations or the initial conditions. 
Currently available tracers of the primordial fluctuations are
the cosmic microwave background (hereafter CMB) anisotropies 
and the large-scale structure (LSS) of the universe. 
From these one can study the geometry of space, matter content,
matter power spectrum (PS), non-Gaussianity of the initial conditions, and so on.
Information from these tracers has limitations because it contains knowledge
only in one thin shell located at a specific epoch in the case of CMB,
or because the amount
of the corresponding data is not yet large enough to constrain cosmological models
strongly in the LSS case. 
The eventual limitation lies in the finite volume of the observable universe.

The second measurable phenomenon of the universe is the expansion of the universe. 
It can be measured by observing the standard candles (e.g. supernova type Ia; Colgate 1979; Riess et al. 1998; Permultter et al. 1999),
the standard rulers (e.g. baryon acoustic oscillations, hereafter BAOs; 
Peebles \& Yu 1970; Meiksin, White,
\& Peacock 1999), 
or standard populations, if any.
Redshifts of these objects give us the relation between the comoving
distance $r$ and redshift $z$ through the luminosity distance $D_L(z)$
and/or angular diameter distance $D_A(z)$,
which constrain
the expansion history of the universe or the Hubble parameter $H(z)$ through the relation
$r(z) = \int_0^z dz'/H(z')$.
The Hubble parameter depends on many cosmological parameters such as the 
total density parameter $\Omega_{\rm tot}$, matter density parameter $\Omega_m$, 
and the equation of state of the dark energy $w=P/\rho$. 
However, there are various kinds of systematic effects that limit the power of
this method. For example, the dependences of the `standard' properties on tracer
subclasses and on redshift are the most serious error sources in measuring
$r(z)$ in the case of the standard candles and populations. The standard rulers
also suffer from all kinds of systematics such as non-linear 
gravitational evolution, redshift-space distortion, past light-cone effects, 
and biasing of tracers. 

The third phenomenon is the growth of cosmic structures, which depends on
both expansion history and initial matter fluctuations. This can be examined
by observing the integrated Sachs-Wolfe effect causing a correlation between
CMB anisotropy and LSS (Sachs \& Wolfe 1967, Corasaniti et al. 2003),
abundance of galaxy clusters (Allen et al. 2004, Rapetti et al. 2005),
and the weak  gravitational lensing by LSS (Cooray \& Huterer 1999).
Properties of some non-linear objects can be also used.
Various present and redshift-dependent properties of intergalactic medium
(near the reionization epoch, in particular), massive dark halos  (luminous galaxies 
and clusters of galaxies), etc., are the combined results of the initial matter 
fluctuations, expansion history, and non-linear physics. 

In this paper we propose to use the pattern of the large-scale galaxy distribution 
to study both the first and second phenomena of the universe. 
We will introduce this tool as a geometrical method similar to the
Alcock-Paczynski test (Alcock \& Paczynski 1979) or 
the BAO-scale method (Blake \& Glazebrook 2003).
In the forthcoming papers we will also show that this method is 
complementary to other methods such as the BAO-scale method, and 
has a power comparable to the BAO method in constraining the dark energy
equation of state (Kim et al. 2009).

\section{Large-scale structure as a standard ruler}

The large-scale distribution of galaxies has long been used to constrain
cosmological models through two-point correlation function (hereafter CF; Davis \& 
Peebles 1983; Maddox et al. 1990) and PS analyses (Park, Gott, \& da Costa 1992; Vogeley
et al. 1992; Park et al. 1994, Tegmark et al. 2006)
because the shapes of the PS and CF depend on the cosmological parameters
such as the matter and baryon density parameters ($\Omega_{m}h^2, 
\Omega_{b}h^2$), and the primordial spectral index ($n_s$). 

Sensitivity to the expansion of the universe appears when
the LSS observation spans a range of redshift. 
Comparing of the shape of the PS at two different epochs, 
knowing they should be the same, one can find how the universe has expanded 
between the epochs. This would have been impossible
if the universe had a scale-free PS since there is no characteristic scale 
to compare the spectra at two epochs. 
But the curvature (the scale-dependence of the slope) in the observed CF (Maddox et al. 1990) 
and the PS (Vogeley et al. 1992; Park et al. 1994) has been well confirmed.
Theoretically, the linear density PS of the Cold Dark Matter models 
is expected to have a peak near the scale corresponding
to the epoch of matter-radiation equality, approaches $k^{n_s}$
at the largest scales, and $k^{-3}$ at the smallest scales.
The PS or CF is only one
of many properties of LSS that can be used as standard rulers, and another
is the topology.
Bond et al. (1996) pointed out that the filament-dominated cosmic web is present
in embryonic form in the overdensity pattern of the initial fluctuations with
non-linear dynamics just sharpening the image. 
However, it is not just overdensity pattern
like the cosmic web structure but the whole large-scale structure 
including the cosmic voids that memorizes their initial
birth places.
After all, it is the whole cosmic sponge rather than just its high density side 
that remains unchanged through the evolution of the universe.
In the case of the flat $\Lambda$CDM model with the WMAP 5-year cosmological
parameters
of $\Omega_m=0.26, \Omega_{\Lambda}=0.74, \Omega_b=0.044, h=0.72, \sigma_8=0.796$,
and $n_s=0.96$ (Dunkley et al. 2009), our N-body simulations show that
the RMS matter displacements till redshifts $z=0.5$ and 0 are 7.7
and 9.7 $h^{-1}$Mpc, respectively.  At the scales much larger than the RMS
displacement the topology of LSS should represent that of the initial density 
fluctuations accurately.

%
%

In this paper we will adopt the genus statistic 
(Gott, Dickinson \& Melott 1986) as a measure of the LSS topology. 
For Gaussian random phase initial conditions 
the genus curve is given by
\begin{equation} 
\label{eq:GRgenus}
g(\nu) = A(1 - \nu^2)e^{-\nu^2/2},
\end{equation}
where $\nu$ is the density threshold level normalized by the RMS density
fluctuation, and the amplitude $A = (\langle k^2 \rangle /3)^{3/2}/2\pi^2$
and $\langle k^2 \rangle$ is the average value of $k^2$ in the
smooth PS (Hamilton, Gott \& Weinberg 1986; Doroshkevich 1970). 
The amplitude $A$ measures the slope of the PS near the smoothing scale, 
and is independent of the amplitude of the PS.  According to most previous 
studies of non-Gaussianity of the primordial density fluctuations, the 
large-scale distribution of galaxies is consistent with the cosmological models 
with initially-Gaussian matter density fields (e.g. Gott et al. 2009). 

The topology of LSS at a given scale is conserved in the course of 
time in the comoving space regardless of whether the initial topology is 
Gaussian or non-Gaussian as long as the scale corresponds to the linear regime. 
When the primordial field is not Gaussian, $A$
is not simply determined by the shape of the PS. However, $A$
is still a conserved quantity and can be used for reconstructing
the expansion history of the universe.
In this sense the topology of LSS is a cosmic standard ruler independent of the PS.

The genus is measured from the iso-density contour surfaces of the smoothed
galaxy distribution. 
Being a measure of intrinsic topology, the genus is insensitive 
to the galaxy biasing and redshift-space distortions.
This is because the intrinsic topology does not change as the shape
of a structure continuously deforms without breaking up or connecting
with itself or other structures.  Furthermore,
according to the second-order perturbation theory, there is no change 
in $A$ due to the weak non-linear 
gravitational evolution 
(Matsubara 1994). 
Namely, the genus amplitude is a powerful measure of the slope of the primordial PS
in the case of a Gaussian field.
It is because of this property of the genus topology that we prefer to adopt it 
as a standard ruler rather than directly using the PS or CF. 

\begin{figure}
\epsscale{1.}
\plotone{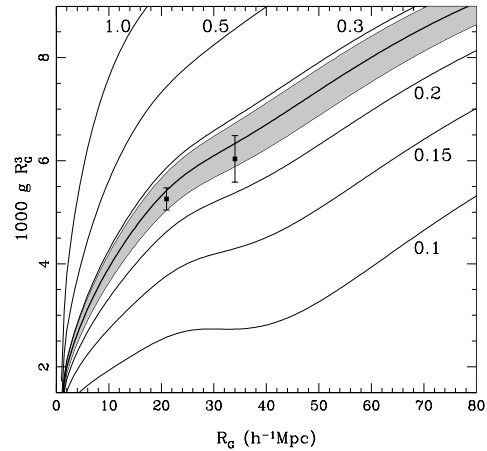}
\caption{
Amplitude of the genus curve per unit smoothing volume
for the flat $\Lambda$CDM models with $\Omega_m+\Omega_{\Lambda}=1$.
The label on each line is $\Omega_m$
and the remaining cosmological parameters are set to those from
the WMAP 5-year data. The thick line is for $\Omega_m=0.26$,
and the boundaries of the shaded region correspond to
$\Omega_m=0.29$ and 0.23, which are chosen from the mean and the $1\sigma$ 
uncertainty limits of the WMAP 5-year data. The data points are from the SDSS
DR4plus sample (Gott et al. 2009).}
\end{figure}

Figure 1 shows 
the amplitude of the genus curve per unit smoothing volume, $gR_{G}^{3}$, 
when the matter field is smoothed over $R_{G}$ by a Gaussian filter, 
for a series of the flat $\Lambda$CDM models with 
$\Omega_{\rm m}+\Omega_{\Lambda}=1$ and $h=0.72$.
The label on each curve is the value of $\Omega_{\rm m}$. 
Each curve has a characteristic shape reflecting the shape of the $\Lambda$CDM 
PS with different $\Omega_{\rm m}$. 
A comparison between these curves and $gR_G^3$ measured from low redshift
LSS data constrains the cosmological parameters related with the shape of the PS.
The uncertainty in $\Omega_m$ from the WMAP
5-year data corresponds to 6.6\% variation in the genus amplitude at
$R_G=20 h^{-1}$Mpc, for example (see the shaded region in Fig. 1).
The two data points are from Gott et al. (2009) who measured $A$ with 4\% 
uncertainty at 21 $h^{-1}$Mpc using the SDSS DR4plus sample.

%

\section{Expansion History of the Universe}

As the universe expands,
the cosmic sponge remains unchanged and the shape of
the PS at large scales is conserved in the comoving space. 
If an observational sample of LSS covers 
a range of redshift, one can measure the PS or CF
in different redshift intervals and compare their shapes to 
check whether or not the adopted $r(z)$ relation is correct. 
If their shapes in different redshift intervals do not agree
with one another, we change the cosmological parameters and thus $r(z)$
relation until an agreement is achieved.
In this section we will consider only the Gaussian fluctuation case,
where the genus amplitude gives 
information equivalent to the shape of the PS,
for a demonstration of concept. 

\begin{figure}
\epsscale{0.7}
\plotone{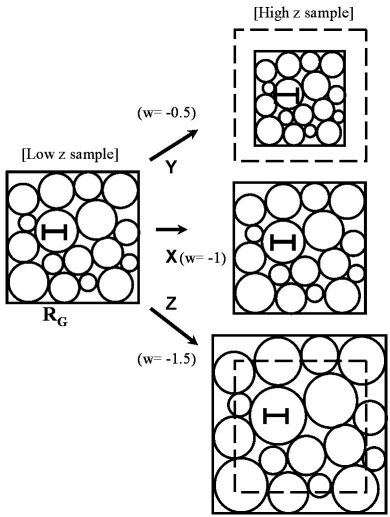}
\caption{
A schematic diagram illustrating changes in the unit volume (dashed boxes) 
and smoothing scale (error bars)
when different $r(z)$ relations are adopted for large-scale structure
data observed at low and high redshifts,
supposing the true cosmology is X and the assumed cosmologies 
are Y and Z with wrong $w$'s.
}
\end{figure}

Figure 2 illustrates what happens to the LSS analysis when wrong
$r(z)$ relations (wrong cosmologies) are adopted. 
Suppose one is trying to constrain 
$w$ while other parameters are fixed. 
The box on the left shows LSS at low redshift in a square region 
of the universe with $w=-1$. 
The error bar indicates the
smoothing scale. If one transforms the redshifts of distant galaxies 
to comoving distances using the $r(z)$ relation adopting the correct 
value of $w$, a region having the same comoving size located
at high redshift will enclose the same comoving 
volume and the smoothing length will correspond to the scale equal to 
that at $z=0$ (the middle box on the right). 
But if we choose $w=-0.5$, we mistakenly think
the space is expanding slower than the reality.
As a result of the wrong $r$-$z$ transformation, a unit volume 
(the dashed box in Fig. 2) will enclose more LSS than the box does at $z=0$, 
and the smoothing length corresponds to a scale larger 
than that at low $z$. Because the box of a unit comoving volume 
contains more LSS but the smoothing is made over a larger scale, 
their effects on the genus 
partially cancel with each other but there remains some net effect
unless the density fluctuation field is scale-free.
When one adopts a universe that expands faster than the real one
($w=-1.5$ for example), the comoving volume of the box at high redshift 
actually amounts to a smaller volume compared to that in the true cosmology, 
and the smoothing scale corresponds to a scale smaller than
what it is intended to.

The amplitude of the genus curve when a wrong cosmology `Y' is adopted
while the true cosmology is `X', can be estimated in the following way.
The volume factor at redshift $z$ in a cosmology is given by 
$V=D_{A}^{2}/H(z)$ (Peebles 1993).
When a wrong cosmology is adopted, the fractional change in volume
is $V_{X}(z)/V_{Y}(z)$ when the samples are constrained to have 
the same comoving volume under the given cosmologies. 
(Due to the Alcock-Paczynski effect, the amount of radial and tangential 
length variations is slightly different from each other.
In the present treatment we average structures over angles 
and consider only the volume effect.)
On the other hand, the smoothing length changes by 
a factor $\lambda_{XY}(z)=[V_{X}(z)/V_{Y}(z)]^{1/3}$. Therefore, the amplitude of 
the genus curve measured when the wrong cosmology $Y$ is adopted
to convert redshifts to comoving distance, becomes
\begin{equation}
g_{\rm Y}(z;R_{G})=g_{\rm X}(z;R_{G}') V_{\rm X}(z)/V_{\rm Y}(z),
\end{equation}
where $R_{G}'=\lambda_{XY}(z) R_{G}$.
This formula is equivalent to $g_{\rm Y}(z;R_{G})R_G^3=g_{\rm X}(z;R_{G}'){R'}_G^3 $.
Any change in cosmology that affects the expansion history of the universe
will result in change in the redshift dependence of $g(R_G)$.
One can use this formula to estimate the genus
in a particular cosmology when its value for a fiducial cosmology is known
without making full numerical mock sample analysis.  Cosmological parameters can be
constrained through an iterative process to minimize the difference between
observations and the theoretical prediction.
In practice, observational data occupy a finite redshift interval, and
the scaling factors in Equation (2) should be replaced by 
an integral over redshift.
It should also be pointed out that the number density of the objects 
tracing the LSS will show a radial gradient due to the
redshift dependence of the volume factor when a wrong cosmology is adopted.
If the radial distribution of the tracers is forced to be uniform,
the selection criterion of the objects will become non-uniform.

When one measures the two-point CF under the assumption of a wrong 
cosmology $Y$, one will incorrectly scale the separation between galaxies,
and the position of features in the CF moves by the scaling factor, namely
$\xi_Y(s;z) = \xi_X(s';z)$ where $s'=\lambda_{XY}(z)s$ 
(see also Percival et al. 2007).
This causes the slope of the CF to change.  Likewise, the PS is scaled as
$P_Y(k;z) = P_X(k';z)$ where $k'=\lambda_{XY}^{-1}(z)k$.

\begin{figure}
\epsscale{1.2}
\plotone{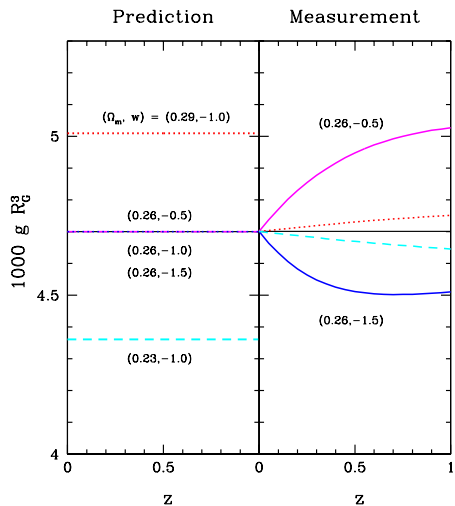}
\caption{
The genus per unit smoothing volume at $R_G=15h^{-1}$Mpc for five flat cosmological
models with different sets of $(\Omega_m,w)$ (left panel). 
The right panel shows the genus that will be actually 
measured when five different sets are assumed in the $r$-$z$ transformation
even though the correct one is $(0.26,-1.0)$.
}
\end{figure}

Figure 3 shows how the genus is used to measure the cosmological 
parameters that govern the expansion history of the universe.
Here it is again assumed that the true cosmology has $\Omega_{\rm m}=0.26$ and
$\Omega_{\rm E}=0.74$ with $w=-1$ (The subscript E stands for the dark energy). 
We adopt $\Omega_m+\Omega_{\rm E}=1$, and all
remaining parameters are fixed to the WMAP 5-year parameters.
The left panel shows the predicted amplitude of the genus curve
at $R_{G}=15\;h^{-1}{\rm Mpc}$ when different sets of ($\Omega_{\rm m},w$) 
are adopted. Each line is independent of $z$
because the shape of the linear PS is conserved in the comoving space.
It is also independent of $w$ since $w$
does not change the shape of the linear PS.
The lines in the right panel are the genus per unit smoothing 
volume that will be measured when five different cosmologies are adopted.
It demonstrates that the redshift-dependence of the genus amplitude is 
quite different for cosmologies with different $\Omega_m$ and $w$.
The correct choice of (0.26, $-1.0$) will 
result in an agreement between the theoretical prediction,
the horizontal line with $gR^3_G=0.00470$, and the observationally measured one.
But when (0.26, $-0.5$) is mistakenly adopted, the wrong $r(z)$ 
makes the genus amplitude overestimated. 
This means the volume factor dominates the change in Equation 2. 
At $z=0.5$, the measured value will be 5.3\% larger than the predicted value. 
When the fractional error in the observed $g$ is $\Delta_g$, the fractional error
in $w$ constrained by this single data point is roughly $\Delta_w = 9.5 \Delta_g$.
Therefore, if the genus is measured with accuracy better than 1\% at $z \sim 0.5$, 
one can constrain $w$ with error less than about $10\%$ by comparing
the theoretical prediction with what is actually measured.
When a model with more negative $w$ is 
adopted, the measured genus amplitude falls below the predicted value
[see the (0.26, $-1.5$) case].

\section{Summary}
The key points of this paper can be summarized as follows:

\begin{enumerate}
 \item We point out that the pattern of the LSS is conserved in time
and can be used as a cosmic standard ruler. 
\item We propose to use the sponge topology of LSS as a 
more robust standard ruler than $P(k)$ or $\xi(r)$ because
the intrinsic topology statistics are less susceptible to various non-linear 
systematic effects.
\end{enumerate}

To enhance the power of the standard rulers 
it is necessary to have accurate knowledge on the systematic effects
such as non-linear gravitational evolution, scale-dependent galaxy biasing,
and redshift-space distortion (Park, Kim \& Gott 2005; James et al. 2009).
This will be the subject of our subsequent paper (Kim et al. 2009).

Recently, Gott et al. (2009) measured the genus amplitude with 
4\% error at the smoothing scale of 21 
$h^{-1}$Mpc from the Luminous Red Galaxy sample of the SDSS DR4plus.
The current redshift survey sample already reached the size that 
enables us to do the topology study at a few percent uncertainty level.
The final SDSS DR7 data and the future LSS surveys 
are expected to significantly increase the accuracy.
Various constrains on $w$ from existing and future surveys will be presented
by Kim et al. (2009)

In addition to the 3D genus, one can also use the scale-dependence 
of the 1D level-crossing statistic or the 2D genus of LSS
as the standard rulers.
For example, cosmological parameters can be estimated by requiring 
the level crossings per unit comoving length in the radial distribution 
of $Ly$-$\alpha$ forest clouds to be constant in time.
The 2D genus of the galaxy distribution in the 
photometric redshift slices can be 
also used to constrain the expansion history of the universe.  
We will explore the usefulness of these statistics 
in the forthcoming papers.

\acknowledgments
CBP acknowledges the support of the Korea Science and Engineering
Foundation (KOSEF) through the Astrophysical Research Center for the
Structure and Evolution of the Cosmos (ARCSEC).

{}

\begin{thebibliography}{}
\bibitem[],{}Alcock, C., \& Paczynski, B. 1979, Nature, 281, 358
\bibitem[],{}Allen, S. W., Schmidt, R. W., Ebeling, H., Fabian, A. C. \& van Speybroeck, L., 2004, MNRAS, 353, 457
\bibitem[],{}Blake, C., \& Glazebrook, K. 2003, ApJ, 594, 665
\bibitem[],{}Bond, J. R., Kofman, L., \& Pogosyan, D., 1996, Nature, 380, 603
\bibitem[],{}Colgate, S., 1979, ApJ, 232, 404
\bibitem[],{}Cooray, A. R., \& Huterer, D. 1999, ApJ, 513, 95
\bibitem[],{}Corasaniti, P. S., Bassett, B. A., Ungarelli, C., \& Copeland, E. J. 2003, Phys.Rev.Lett. 90, 091303
\bibitem[],{}Doroshkevich, G. 1970, Astrophysika, 6, 320
\bibitem[],{}Dunkley, J., Komatsu, E., Nolta, M. R., Spergel, D. N., Larson, D., Hinshaw, G., Page, L., Bennett, C. L., et al. 2009, ApJS, 180, 306
\bibitem[],{}Gott, J. R., Choi, Y.-Y., Park, C., \& Kim, J. 2009,ApJ, 695, L45
\bibitem[],{}Gott, J. R., Melott, A. L., \& Dickinson, M. 1986, \apj, 306, 341
\bibitem[],{}Hamilton, A. J. S., Gott, J. R., \& Weinberg, D. W. 1986, \apj, 309, 1
\bibitem[],{}James, J. B., Colless, M., Lewis, G. F., \& Peacock, J. A. 2009, MNRAS, 394, 454
\bibitem[],{}Kim, Y.-R., Park, C., Kim, J., \& Gott, J., R., 2009, in preparation
\bibitem[],{}Maddox, S. J., Sutherland, W. J., Efstathiou, G., \& Loveday, J.
1991, in Large-Scale Structures and Peculiar Motions in the Universe, ASP
Conference Series, Vol. 15, ed. D. W. Latham, \& L. N. da Costa, p. 213
\bibitem[],{}Meiksin, A., White, M., \& Peacock, J. A. 1999, MNRAS, 304, 851
\bibitem[],{}Park, C., Choi, Y.-Y., Vogeley, M. S., Gott, J. R., Kim, J., 
Hikage, C., Matsubara, T., Park, M. G., Suto, Y., \& Weinberg, D. H.
2005$b$, \apj, 633, 11
\bibitem[],{}Park, C., Kim,  J., \& Gott, J. R. 2005, \apj, 633, 1
\bibitem[Peebles(1993)]{pee93} Peebles, P.~J.~E.\ 1993, 
Princeton Series in Physics, Princeton, NJ: Princeton University Press, p332
\bibitem[]{}Peebles, P.~J.~E., \& Yu, J. T. 1970, ApJ, 162, 815
\bibitem[],{}Percival, W. J., Cole, S, Eisenstein, D. J., Nicho, R. C., Peacock, J. A., Pope, A. C., \& Szalay, A. S., 2007, MNRAS, 381, 1053
\bibitem[],{}Perlmutter, S., Aldering, G., Goldhaber, G., Knop, R. A., Nugent, P., Castro, P. G., Deustua, S., Fabbro, S., et al., 1999, \apj, 517, 565
\bibitem[],{}Rapetti, D., Allen, S. W. \& Weller, Jochen, 2005, MNRAS, 360, 555
\bibitem[],{}Riess, A. G., Filippenko, A. V., Challis, p., Clocchiatti, A., Diercks, A., Garnavich, P. M., Gilliland, R. L., Hogan, C. J, et al. 1998, \aj, 116, 1009
\bibitem[],{}Sachs, R.~K., \& Wolfe, A. M., 1967, ApJ, 147, 73
\bibitem[],{}Tegmark, M., Eisenstein, D. J., Strauss, M. A., Weinberg, D. H., Blanton, M. R., Frieman, J. A., Fukugita, M., Gunn, J. E., et al. 2006, Phys. Rev. D., 74, 123507
\bibitem[],{}Vogeley, M. S., Park, C., Geller, M. J., Huchra, J. P. \& 
Gott, J. R. 1994, \apj, 420, 525
\end{thebibliography}
\end{document}